\documentclass[12pt]{article}
\usepackage[a4paper, total={21cm, 14cm}, margin=2.5cm]{geometry}
\usepackage{setspace}
\usepackage{amsmath} 
\usepackage{array} 
\usepackage{graphicx} 
\usepackage{xcolor}
\usepackage{hyperref} 
\bibliography{INJP}

\begin{document}
	
	\title{Heavy quarkonium dissociation in the presence of magnetic field and anisotropy using dissociation energy criterion}
	\author{
		Rishabh Sharma, Siddhartha Solanki, Manohar Lal and Vineet Kumar Agotiya\thanks{Email: agotiya@gmail.com} \\
	Central University of Jharkhand, Ranchi, India
	}
	\date{\today}
	
	\maketitle
		\begin{abstract}
				
		In this article, we have studied the dissociation temperature of $1S$ and $2S$ states of heavy quarkonium in the presence of anisotropy and a strong magnetic field background using the dissociation energy criterion. We utilized the medium-modified form of the Cornell potential, which depends on temperature as well as the anisotropic parameter $\xi$ and the magnetic field. The binding energy (B.E.) and dissociation energy (D.E.) of heavy quarkonium have been examined for different values of the magnetic field and anisotropy. It is noted that B.E. starts decreasing from higher values as we increase the anisotropy, while D.E. exhibits the opposite behavior. The dissociation temperature appears to increase with anisotropy, while it decreases with the magnetic field, as shown in Table \ref{T1} and \ref{T2} respectively. These results align well with recent research findings.
		\vspace{1em}
		
	
	\end{abstract}

\maketitle

\section{\label{sec:level1}Introduction}

The primary objective of heavy ion collider experiments, such as the Relativistic Heavy Ion Collider (RHIC) at Brookhaven National Laboratory (BNL) and the Large Hadron Collider (LHC) at European Organization for Nuclear Research (CERN), is to create a small volume of matter and heat it to extreme temperatures to achieve the deconfinement of quarks and gluons. Lattice Quantum Chromodynamics (QCD) indicated that the crossover transition of strongly interacting matter~\cite{Petreczky,Bazavov_2009,Bazavov_2010,Borsányi,Cheng} to deconfined states of quarks and gluons occur at approximately 130-200 MeV, equivalent to a temperature of about 2 trillion degrees Kelvin. In Au-Au collisions at RHIC, the center-of-mass energy at the time of collision was measured at $\sqrt{S_{NN}}$=200 GeV, with a maximum central temperature of about $T_o$= 450 MeV. In the current LHC run, a center-of-mass energy of $\sqrt{S_{NN}}$= 2.76 TeV was achieved~\cite{Schenke}, resulting in a central temperature of 550 MeV. For the upcoming LHC run, operations are planned at 5.5 TeV, with an estimated temperature of 700-800 MeV.
\\
The ultra-relativistic heavy ion collision experiments conducted at RHIC and LHC provide crucial insights into the behavior of Quark-Gluon Plasma (QGP), demonstrating that it behaves like a perfect fluid rather than a non-interacting gas of heavy quarks and anti-quarks~\cite{Aamodt_2010,Aamodt_2011,Adcox_2005,Back_2005,Arsene_2005}. Experimental observations indicate that quarkonium suppression offers compelling evidence for the formation of QGP during heavy ion collisions~\cite{Aamodt_2011,Adams_2005,Adcox_2005,Back_2005,Arsene_2005,McLerran}. This suppression highlights the plasma characteristics of the medium, including phenomena such as color screening~\cite{Chu_1989}, Landau damping~\cite{Landau_1984}, and energy loss~\cite{Koike}. Since the discovery of $J/\psi$ in 1974~\cite{Augustin_1974,Aubert} both theoretical and experimental studies of quarkonium have continued to be a significant area of research.\\
The dissociation of quarkonium—a color-singlet state composed of quarks and their corresponding anti-quarks, bound by static, charge-neutral gluons—in deconfined states at finite temperatures due to color screening was first studied by Matsui and Satz~\cite{Matsui_1986}. Since then, numerous articles have been published, providing important refinements in the study of heavy quarkonia~\cite{Jamal_2018,Patra_2010,Brambilla_2018}.

Beyond being a bound state of quarks and anti-quarks, quarkonium plays a crucial role in understanding significant aspects of Quantum Chromodynamics (QCD). In recent years, extensive studies have investigated the properties of heavy quarkonia~\cite{Luchinsky,Fulsom}, including examinations of the dissociation temperature ~\cite{Brambilla_2005,Brambilla_2013,Braga,Duclou,Chen_2017}. At high temperatures, strongly interacting matter transitions into a deconfined state of quarks and gluons, which is characterized as a crossover transition rather than a phase transition. This led to the emergence of the concept of Debye screening (the separation between quarks), facilitating a better understanding of the dissociation of hadronic states~\cite{Shuryak}.

The spectrum of bound states in hadronic matter can be effectively described by the Non-Relativistic Potential Model~\cite{Matsui_1986,Karsch}. The binding energy of heavy quarkonium is significantly lower than the quark mass ($m_q$), and its size is much larger than 1/$m_q$. Effective field theory of QCD provides a framework for describing non-relativistic heavy quarkonium states. Given that the velocity of quarks at zero temperature is much less than the speed of light, quarkonia are studied using the non-relativistic potential model, from which the Cornell potential can be derived directly~\cite{Brambilla_2005,Lucha,Eichten_1980}.   
Using the Non-Relativistic Potential Model, one can readily determine the spectral function~\cite{Mocsy_2005,Mocsy_2006,Cabrera,Alberico,Mocsy_2008}. In recent years, a new theoretical development has emerged regarding thermal width, which arises from the imaginary component of the potential. For the first time, Laine et al.\cite{Laine_2007,Laine_HEP} derived the imaginary part of the potential using gluonic Landau damping within the framework of leading-order perturbative theory.

The calculation of the potential is extended to include an anisotropic factor associated with quarkonia at all stages. The contributions of anisotropy are evaluated for both the real part~\cite{Dumitru_PLB,Dumitru_PRD} and the imaginary part~\cite{Burnier,Philipsen} of the potential. The effects of local momentum anisotropy have been examined in several studies~\cite{Martinez_2011,Martinez_2010,Ryblewski,Florkowski,Muronga}. Numerous articles investigate the properties of quarkonium under the influence of anisotropic parameters~\cite{Dumitru_PLB,Dumitru_PRC,Martinez_PRL,Martinez_PRC,Baier} and strong magnetic field backgrounds. It is important to note that Quark-Gluon Plasma (QGP) is never found in an isotropic form; rather, it consistently exhibits significant momentum anisotropy, this motivates us to considers the effect of momentum space anisotropy along with the effect of temperature and magnetic field.

In this manuscript, we account for the presence of anisotropy and magnetic fields to estimate the dissociation temperatures of the 1S and 2S states of heavy quarkonia. We employ a novel potential model method to examine the dissociation temperature using the dissociation energy criterion. Our findings for dissociation temperature values, based on the dissociation energy criterion, are approximately aligned with previously published results~ \cite{Solanki_PRC} obtained using the thermal width and thermal energy criteria. The study of heavy quarkonia under the influence of varying parameters, such as anisotropy and magnetic fields, offers valuable insights into their dynamics in extreme conditions. By investigating the effects of anisotropy, we enhance our understanding of the potential energy landscape of quarkonium.

\section{\label{sec:level2}Model Setup}
\subsection{\label{subsec:level1}Debye screening mass in the presence of magnetic field}
The Debye mass $m_D$ is a crucial parameter for understanding the screening of color forces in a hot QCD medium. It is known that plasma exhibits collective behavior due to the presence of both charged and neutral quasi-particles. The Debye mass can be related to the screening length (the inverse of $m_D$) because of its ability to shield electric potentials applied to it. A detailed description of the Debye mass is provided in the literature~\cite{Rebhan,Yagi}.\\
In the quasi-particle framework, all interacting particles are generally treated as non-interacting. This description can also be understood through effective fugacity~\cite{Chandra_2009} or effective mass~\cite{Goloviznin,Peshier}. Various models, such as the Nambu-Jona-Lasinio (NJL) model, the Polyakov-Nambu-Jona-Lasinio (PNJL) model~\cite{Dumitru_2002}, and self-consistent quasi-particle models~\cite{Bannur_2006,Bannur_2007}, explain effective masses in detail.\\
In this manuscript, we employ the Effective Quasi-Particle Model (EQPM) in the presence of a magnetic field background, which describes the QCD equation of state (EOS) in terms of effective fugacity parameters $z_q$ for quarks and anti-quarks, and $z_g$ for gluons. We also explore the distribution functions for quasi-gluons and quasi-quarks/antiquarks~\cite{Chandra_2012,Kurian_2017,Kurian_2018} is 
\begin{equation}
	\label{1}
	f^0_{g/q}= \frac{f_{g/q}e^-\beta E_p}{1 \mp f_{g/q}e^-\beta E_p}
	 \end{equation}
Since in high energy physics $\hbar$ = $c$ = $k_B$ = 1, $\beta=1/T$ and $E_{p}$ represents the Landau energy eigen value for quasiquarks/anti-quark and for quasi gluons:
\begin{equation}
	\label{2}
	E_{p} = \sqrt{p^2 + m_q^2+ 2l|q_f eB|}  \quad \quad  \text{(for quasiquarks/antiquark)}   
\end{equation}
In eq. \ref{2}, $`m_q$' refers to the mass of the heavy quark, with values chosen for $J/\Psi$ = 1.5 GeV and $\Upsilon$ = 4.5 GeV. The quantum number $`l$' (where $`l$' = 0,1,2.....) represents the Landau quantum number, and $q_f$ denotes the fractional charge of the quark system. The term $`eB$' indicates the magnetic field, where (B = B$\hat{z}$), and its effects are considered in the direction of the z-axis.

Since the quasi-gluons remain intact in the presence of a strong magnetic field background, so the Landau energy eigenvalue for the quasi-gluon system can be represented as follows:
\begin{equation}
	\label{3}
   E_{p} = |\vec{p}| = p  \quad \quad \text{(for quasi gluons)}  
\end{equation}

substituting eq. \ref{2} in eq. \ref{1} we get,
\begin{equation}
	\label{4}
	f^0_{g/q}= \frac{f_{g/q}e^-\beta \sqrt{p^2 + m_q^2+ 2l|q_f eB|} }{1 \mp f_{g/q}e^-\beta \sqrt{p^2 + m_q^2+ 2l|q_f eB|}}
\end{equation}
 To find out the Debye mass which contains magnetic field, we start from gluon self energy \cite{Schneider_2002} as below:

\begin{equation}
	\label{5}
	m^2_D = \pi_{oo}(\omega=p) \vec{p}\rightarrow 0  
\end{equation}
According to \cite{Nilima}, gluon self-energy modified in terms of the distribution function is given below,
\begin{equation}
\label{6}
m^2_D = \frac{g^2|eB|}{2\pi^2 T}\int_{0}^{\infty} dpf^0_{q}(1-f^0_{q})  
\end{equation}

The above eq. \ref{6}, defining Debye mass, obtained by using the quasi-particle approach. The kinetic theory approach can also serve as an alternative method to obtain the Debye mass. Both the quasi-particle approach and the kinetic theory approach \textcolor{blue}{yields} nearly identical results. The Debye mass for $N_c$ = 3 and
 $N_f$ = 3 will be:
 
\begin{equation}
\label{7}
m_D^2=4\pi\alpha_s\left(\frac{6T^2}{\pi}Polylog\left[2,z_g\right]+\frac{3eB}{\pi}\frac{z_g}{1+z_g}\right)     
\end{equation}

Here $z_g$ is quasi-gluon effective fugacity parameter, for ideal case of equation of states $z_g$ = 1. The Polylog function used in the above expression is written as,
\begin{equation*}
		\label{}
		Polylog[2,z_g]=\sum_{k=1}^{\infty}\frac{z^k}{k^2}
\end{equation*}
 Then, Debye mass in terms of temperature and magnetic field representing non-interacting quarks and gluons is given as\cite{Solanki_2022,Lal_2022,Rishabh24}
\begin{equation}
\label{8}
m_D^2\left(T,eB\right)=4\pi\alpha_s\left(T^2+\frac{3eB}{2\pi^2}\right)	
\end{equation}

Here $\alpha_{s}$(T) is two loop running coupling whose value is as follows;

\begin{equation}
	\label{10}
	\alpha_s\left(T\right)=\frac{6\pi}{\left(33-2N_f\right)\theta}\left(1-\frac{3\left(153-19N_f\right)}{\left(33-2N_f\right)^2}\frac{log2\theta}{\theta}\right)
\end{equation}
Where, 
\begin{equation}
	\label{11}
	\theta=log\left(\frac{T}{\lambda_T}\right)
\end{equation}

$\lambda_T$ is QCD coupling scale and its value is 197 MeV.


\subsection{\label{subsec:level2}Real part of Medium Modified Cornell (MMC) potential in the presence of Anisotropy}
The interaction potential of quarkonium states is characterized by the Cornell potential \cite{Eichten_1978,Sebastian}. This potential includes both Coulombic and string terms, which are responsible for the fundamental features of quantum chromodynamics (QCD)—specifically, the interactions at small distances and confinement at large distances. The Cornell potential is represented as: 
\begin{equation}
	\label{9}
	V(r)= -\frac{\alpha}{r}+\sigma r 
\end{equation}
Here `r' is the effective radius of the heavy quarkonium system, `$\sigma$' is string tension whose value ($\sigma$ = 0.224 $GeV^2$).
\\
Since heavy ion collisions are inherently non-central, thus spatial anisotropy is generated at very early stages. As the quark-gluon plasma (QGP) expands over time, various pressure gradients emerges, which are responsible for spatial and momentum anisotropy factors. In this manuscript, we introduce anisotropy at the level of the particle phase space distribution~\cite{Romatschke_2003,Carrington_2014,Jamal_2017}. We utilize an anisotropic distribution function obtained through re-scaling (stretching and squeezing) and can be expressed as:
\begin{equation}
	\label{12}
    f(p) \rightarrow f_{\xi}(t)= C_{\xi}f(\sqrt{p^2+\xi(p.\hat{n})^2})
\end{equation}
f(p) represents the effective fugacity quasi particle isotropic distribution function~\cite{Chandra_2011} of the particle and $\hat{n}$ indicates the direction of anisotropy in momentum space. The  Parameter $\xi$ indicates the anisotropy of a medium. It is defined as follows: for an isotropic medium, $\xi$ = 0; for an oblate form, $\xi > 0$; and for a prolate form, $ -1 < \xi  < 0 $. Various equations of state (EoS) influence the Debye mass ($m_D$). $C_\xi$ is normalization constant~\cite{Carrington_2014} which is used to make the Debye mass intact from the effect of anisotropy, the value of the normalization constant is proposed as  

\begin{figure}[h]
	\centering
	\includegraphics[width=\textwidth]{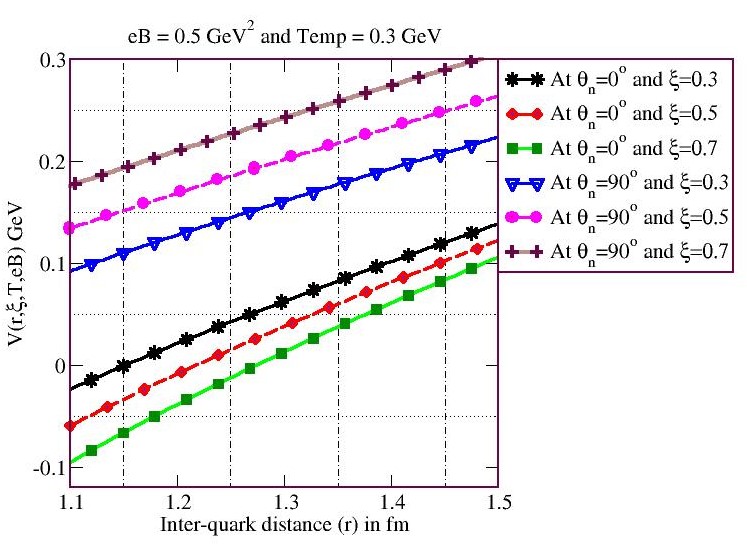} 
	\caption{Shows the variation of heavy quark Potential with inter-quark distance (r) for different values of anisotropy ($\xi$= 0.3, 0.5, 0.7) at $\theta_n$=$0^o$ and $\theta_n$=$90^o$ with fixed value of magnetic field eB = 0.5 $GeV^2$ and Temp = 0.3 GeV.}
	\label{fig.1}
\end{figure}


$$
C_\xi= \begin{cases} 
1-\xi + O(\xi^{3/2}) & $if$ -1\le\xi < 0 $ for prolate case$ \\
1+\xi + O(\xi^{3/2}) & $if$  \xi > 0  $ for oblate case$ \\
\end{cases}
$$

To incorporate the effect of the magnetic field, the Cornell potential has been modified using the Fourier transform. After dividing it with dielectric permittivity we get,

\begin{equation}
\grave{V}(r)= \frac{\bar{V}(k)}{\epsilon(k)}
\label{13}
\end{equation}

To obtain the medium modified form of heavy quark potential, it is essential to calculate the Fourier transform of $\bar{V}(k)$ \cite{Agotiya_2009} and dielectric permittivity ($\epsilon$(k)).
\begin{equation}
	\label{14}
	\bar{V}(k)= -\sqrt{\frac{2}{\pi}}\left(\frac{\alpha}{k^2}+2\frac{\sigma}{k^4}\right)
\end{equation}

Dielectric permittivity $\epsilon$(k) can be accomplished through two methods: (1) by employing gluon self-energy in finite temperature QCD~\cite{Braaten_1990}, and (2) through semi-classical transport theory ~\cite{Blaizot_1994}. Both methods yield the gluon self-energy~\cite{Schneider_2002} ($\Pi^{\mu\nu}$) which provides the static gluon propagator as follows:

\begin{equation}
	\label{15}
	\Delta^{\mu\nu} = k^2 g^{\mu\nu} - k^{\mu} k^{\nu} + \Pi^{\mu\nu}(\omega, k)
\end{equation}
The dielectric tensor of gluon propagator in Fourier space  can be written as:
 
 \begin{equation}
 	\label{16}
 	\epsilon^{-1}(k) = -\lim_{\omega \to 0} k^2 \Delta^{00}(\omega, k)
 \end{equation}
 
 Where $\Delta^{00}$ represents the static limit of ``00" component of gluon propagator. According to linear response theory eq. \ref{16} establishes a connection between dielectric permittivity and the static limit of the gluon propagator.The real part of the dielectric permittivity can be derived from a symmetric operator, expressed as follows:
\begin{equation}
	\begin{split}
			\label{17}
		&\epsilon^{-1}(k)= \pi T m_D^2\bigg(\frac{k^2}{k(k^2+m_D^2)^2}-\xi k^2\bigg(\frac{-1}{3k(k^2+m_D^2)^2}
		+\frac{3 \sin^2{\theta_n}}{4 k(k^2+m_D^2)^2}-
		\frac{2 m_D^2(3 \sin^2{\theta_n}-1)}{3k(k^2+m_D^2)^3}\bigg)\bigg)
	\end{split}
\end{equation}

\begin{figure}[h]
	\centering
	\includegraphics[width=\textwidth]{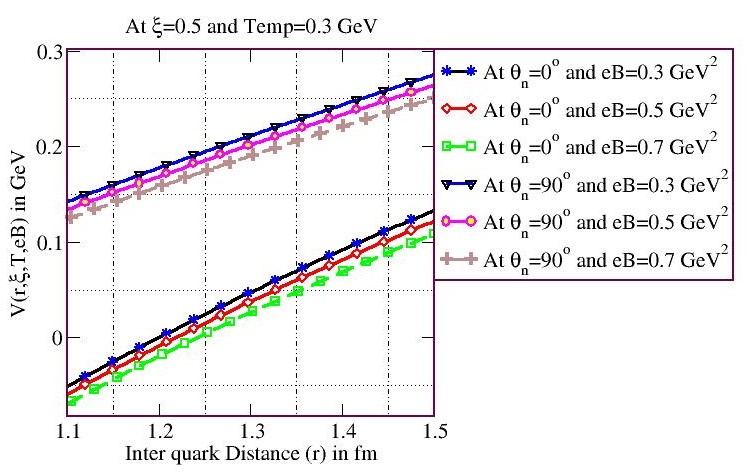} 
	\caption{Shows the variation of heavy quark Potential with inter-quark distance (r) for different values of magnetic fields (eB = 0.3, 0.5, 0.7) $GeV^2$, at $\theta_n$=$0^o$ and $\theta_n$=$90^o$ with a fixed value of anisotropy $\xi$ = 0.5 and Temp = 0.3 GeV.}
	\label{fig.2}
\end{figure}
where,
\begin{equation}
	\label{18}
	\cos{\theta_n}=\cos{\theta_r}\cos{\theta_{pr}}+\sin{\theta_r}\sin{\theta_{pr}}\cos{\phi_{pr}}
\end{equation}
Here $\theta_n$ denotes the angle between particle momentum (p) and anisotropy direction $\hat{n}$, while $\theta_r$  is angle between `r' and $\hat{n}$. $\theta_pr$ and $\phi_pr$ represents the polar and azimuthal angles respectively. Finally, by substituting the  fourier transform of $\bar{V}(k)$ from eq. \ref{14} and $\epsilon^{-1}(k)$ from eq. \ref{17} in eq. \ref{13} we get the real part of heavy quark interaction is obtained as:

\begin{equation}
	\begin{split}
	\label{19}
	Re[V(r,\xi,T,eB)]=\frac{S\sigma}{m_D}\bigg(1+\frac{\xi}{3}\bigg)
	-\frac{\alpha m_D}{S}\bigg(1+\frac{S^2}{2}+\xi\bigg(\frac{1}{3}+\frac{S^2}{16}+\cos(2\theta_r)\bigg)\bigg)
	\end{split}    
\end{equation}

Here $S= rm_D$ and $m_D$(T, eB) represents the magnetic field influenced Quasi particle Debye mass, which is explained in detail in the previous subsection \ref{subsec:level1}.

\subsection{\label{subsec:level3}Binding Energy (B.E.) of 1S and 2S states of heavy quarkonium}

To obtain the results for binding energy as influenced by the anisotropic factor, Schrödinger's equation must be solved for the real part of the heavy quark potential, as done in Refs.\cite{Margotta,Strickland,Thakur_2014}. A key point to note is that the magnitude of the real part of the heavy quark potential is significantly greater than that of the imaginary part. Therefore, The real part of the binding energy can be derived from the radial part of the Schrödinger equation for the isotropic component as well as for anisotropic component is represented as follows:
\begin{equation}
	\begin{split}
	\label{20}
	Re[E_B(\xi,T,eB)]= \bigg(\frac{m_q \sigma^2}{m_D^4 n^2}+ \alpha m_D
	+\frac{\xi}{3}\bigg(3\frac{m_q \sigma^2}{m_D^4 n^2}+\alpha m_D\bigg)\bigg)
\end{split}
\end{equation}
Here $m_q$ represents heavy quark masses. In the present work we used $m_q$  for $J/\Psi$= 1.5 GeV and $\Upsilon$= 4.5 GeV to examine the quarkonium properties taken from\cite{Solanki_PRC}.

\begin{figure*}
	\centering   
	\includegraphics[height=10cm,width=10cm]{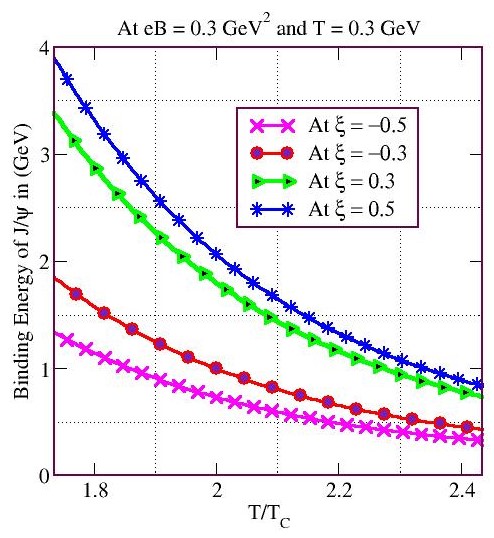}	
	\caption{Shows the variation of B.E. of J/$\psi$ with T/$T_c$ at different values of anisotropy from $\xi$= -0.5 to $\xi$ = 0.5 for fixed values magnetic field of eB = 0.3 $GeV^2$ and Temperature (T= 0.3 GeV)}
	\label{fig.3} 
\end{figure*}
\subsection{\label{subsec:level4}Dissociation energy (D.E.)}

The dissociation energy (D.E.) of heavy quarkonia\cite{Solanki_AHEP} is the most important quantity to detect the vanishing of bound states. Its calculation can be performed using the following relationship;
\begin{equation}
	\label{21}
	D.E. = 2m_q+\frac{2\sigma}{m_D}- B.E.
\end{equation}
By introducing the value of B.E. in the above expression, we get the final expression of D.E.

\begin{equation}
	\begin{split}	
	\label{22}
	D.E.= 2m_q+\frac{2\sigma}{m_D}-\bigg(\frac{m_q \sigma^2}{m_D^4 n^2}+ \alpha m_D+
    \frac{\xi}{3}\left(3\frac{m_q \sigma^2}{m_D^4 n^2}+\alpha m_D\right)\bigg)
\end{split}
\end{equation}
\begin{figure*}
   	\centering
	\vspace{0mm}   
	\includegraphics[height=10cm,width=10cm]{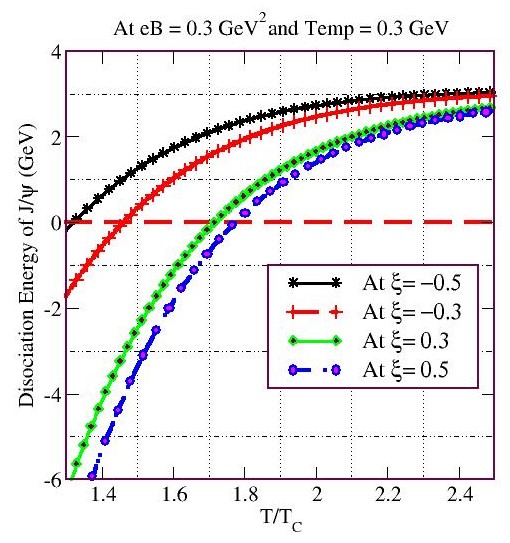}	
	\caption{Shows the variation of D.E. of J/$\psi$ with T/$T_c$ at different values of anisotropy from $\xi$= -0.5 to $\xi$ = 0.5 for fixed value magnetic field (eB = 0.3 $GeV^2$) and Temperature (T= 0.3 GeV).}
	\label{fig.4}
	\vspace{1mm} 
\end{figure*}
\begin{figure*}[ht]
	\vspace{0mm}   
	\includegraphics[height=8cm,width=8cm]{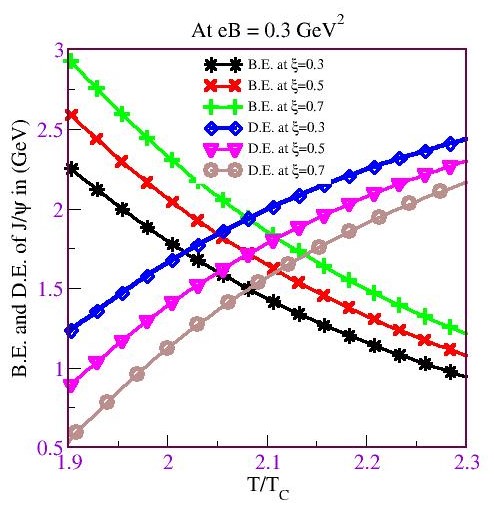}
	\hspace{0mm}
	\includegraphics[height=8cm,width=8cm]{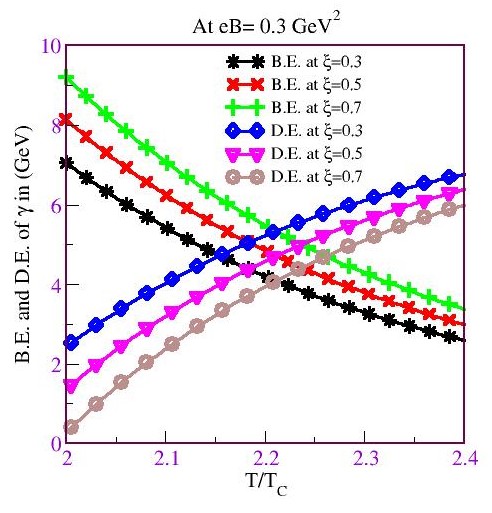}
	\caption{Shows the variation of B.E. and D.E. of $J/\psi$  (left panel) and $\Upsilon$ (right panel) with T/$T_c$ for different values of anisotropy ($\xi$= 0.3, 0.5, 0.7) at a fixed value of magnetic field (eB= 0.3 $GeV^2$)}
	\label{fig.5}
	\vspace{1mm} 
\end{figure*}
\begin{figure*} 
	\vspace{0mm}   
	\includegraphics[height=8cm,width=8cm]{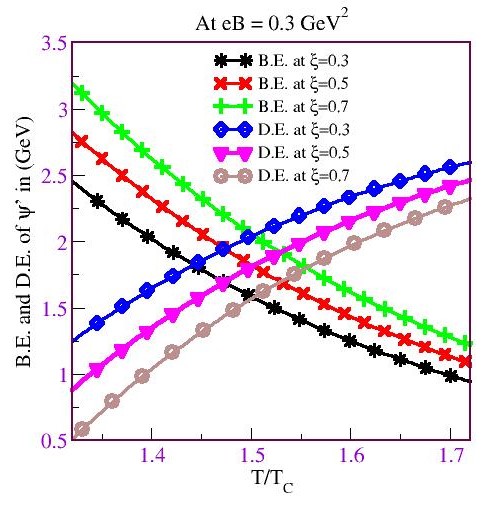}
	\hspace{1mm}
	\includegraphics[height=8cm,width=8cm]{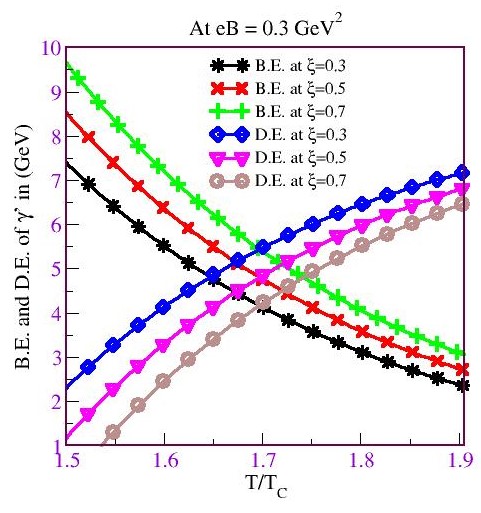}
	\caption{Shows the variation of B.E. and D.E. of $\psi$' (left panel) and $\Upsilon$' (right panel) with T/$T_c$ for different values of anisotropy ($\xi$= 0.3, 0.5, 0.7) at a fixed value of magnetic field (eB= 0.3 $GeV^2$)}
	\label{fig.6}
	\vspace{1mm} 
\end{figure*}
\begin{figure*}
	\vspace{0mm}   
	\includegraphics[height=8cm,width=8cm]{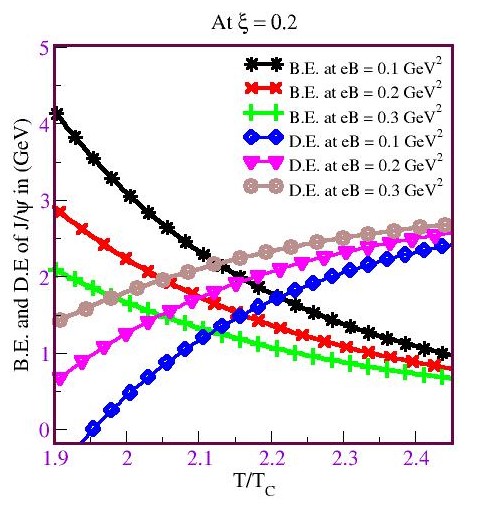}
	\hspace{1mm}
	\includegraphics[height=8cm,width=8cm]{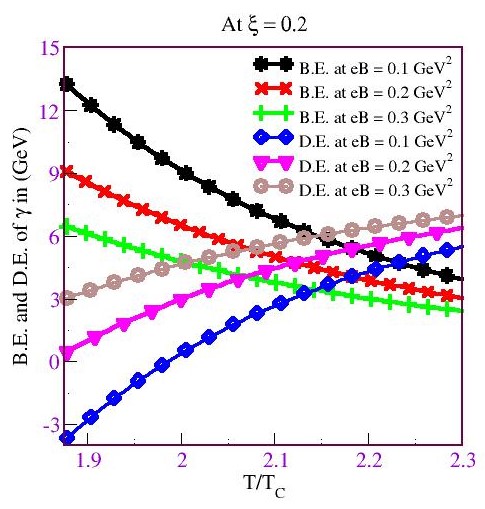}
	\caption{Shows the variation of B.E. and D.E. of $J/\psi$ (left panel) and $\Upsilon$ (right panel) with T/$T_c$ for different values of magnetic field (eB = 0.1, 0.2, 0.3 $GeV^2$) at fixed value of anisotropy ($\xi$= 0.2 )}
	\label{fig.7}
	\vspace{1mm} 
\end{figure*}
\begin{figure*}
	\vspace{0mm}   
	\includegraphics[height=8cm,width=8cm]{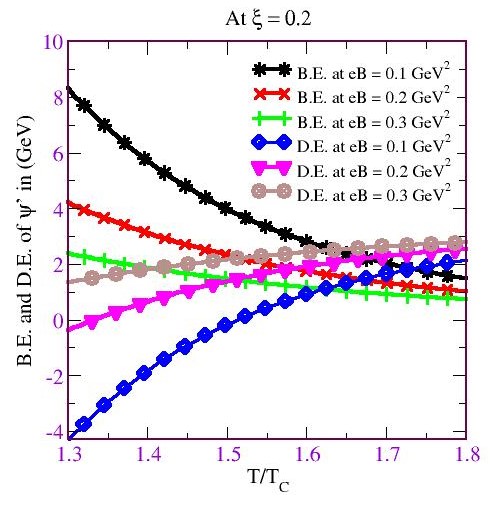}
	\hspace{1mm}
	\includegraphics[height=8cm,width=8cm]{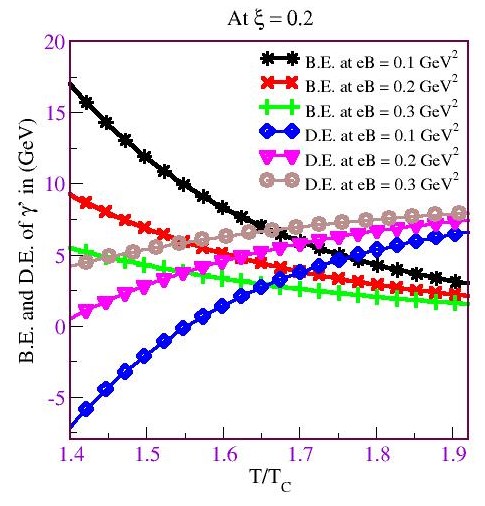}
	\caption{Shows the variation of B.E. and D.E. of $\psi$' (left panel) and $\Upsilon$' (right panel) with T/$T_c$ for different values of magnetic field (eB = 0.1, 0.2, 0.3 $GeV^2$) at fixed value of anisotropy ($\xi$= 0.2 )}
	\label{fig.8}
	\vspace{1mm} 
\end{figure*}

\begin{table}[h]
	\centering
	\begin{tabular}{|c|c|c|c|c|c|}
		\hline
		$\mbox{\boldmath$States\downarrow$}$& *Lower bound\cite{Lal_2023}  & $\mbox{$\xi= 0.3 $}$ & *Upper Bound \cite{Lal_2023} & $\mbox{$\xi= 0.5 $}$ & $\mbox{$\xi= 0.7 $}$\\
		\hline
		$\mbox{\boldmath$J/\psi$}$  & 1.6497  & 2.01 & 2.1827 & 2.08 & 2.13 \\
		\hline
		$\mbox{\boldmath$\psi'$}$ & 1.1421  & 1.44 & 1.5355 & 1.50 & 1.55 \\
		\hline
		$\mbox{\boldmath$\Upsilon$}$ & 2.1700 & 2.15 & 2.8299  & 2.21 & 2.26\\
		\hline
			$\mbox{\boldmath$\Upsilon'$}$ & 1.5355 & 1.64 & 2.030  & 1.69 & 1.74\\
		\hline
	\end{tabular}
	\caption{Dissociation temperature of 1S and 2S states of quarkonium for different anisotropy at fix value of magnetic field eB= 0.3 $GeV^2$.\\ *Column 2 and 4 show the values of lower and upper bounds of dissociation temperature from ref \cite{Lal_2023} at condition $\xi$= 0.3 and eB= 0.3 $GeV^2$.}
	\label{T1}
\end{table}

\begin{table}[h]
	\centering
	\begin{tabular}{|c|c|c|c|}
		\hline
		\textrm{$\mbox{\boldmath$States\downarrow$}$}&
		\textrm{$\mbox{$eB= 0.1 GeV^2$}$}&
		\multicolumn{1}{c}{\textrm{$\mbox{$eB= 0.2 GeV^2$}$}}&
		\textrm{$\mbox{$eB= 0.3 GeV^2$}$}\\
		\hline
		$\mbox{\boldmath$J/\psi$}$ & 2.21 & 2.10 & 1.97\\
		\hline
		$\mbox{\boldmath$\psi'$}$& 1.73 & 1.58 & 1.40\\
		\hline
			$\mbox{\boldmath$\Upsilon$}$& 2.23 & 2.12 & 2.00\\
		\hline
		$\mbox{\boldmath$\Upsilon'$}$& 1.76 & 1.61 & 1.45\\
		\hline
	\end{tabular}
	\caption{Dissociation temperature of 1S and 2S states of quarkonium at fix value of anisotropy $\xi$= 0.2 and temperature for different values of magnetic field.}
	\label{T2}
\end{table}


\section{\label{sec:level3}Results and Discussion}
In the analysis, we have fixed the critical temperature at ($T_c=197 MeV$) throughout the manuscript. We studied the variation of binding energy and dissociation energy of the 1S and 2S states of quarkonium under different values of magnetic field and anisotropy. The variation of heavy quark potential (Eq. \ref{19}) with inter quark distance `r' (in fm) for different values of anisotropy ($\xi$= 0.3, 0.5, 0.7) at fixed value of temperature (T= 0.3 GeV) and magnetic field (eB= 0.5 $GeV^2$) is shown in Fig. \ref{fig.1}. In Fig. \ref{fig.1} when we increase the value of $\xi$ from 0.3 to 0.5 for $\theta$=$0^o$ the value of potential decreases whereas for $\theta$=$90^o$ the variation of the potential increases with increasing anisotropy from 0.3 to 0.5.\\
Fig. \ref{fig.2} also shows the variation of potential with inter quark distance at fixed anisotropy ($\xi$=0.5) and temperature (T= 0.3 GeV), while magnetic field vary from 0.3 $GeV^2$ to 0.7 $GeV^2$. From Fig. \ref{fig.2} we see that for both the cases ($\theta$=$0^o$ and $\theta$=$90^o$), the value of potential decreases by increasing the value of the magnetic field.\\
Fig. \ref{fig.3} shows the variation of B.E. of J/$\psi$ with temperature. From Fig. \ref{fig.3}, it is seen that B.E. of J/$\psi$ decreases with temperature, for different values of anisotrpy. On increasing the value of anisotropy parameter from ($\xi$= -0.5, -0.3, 0.3, 0.5) B.E. is increases at fix value of magnetic field (eB= 0.3 $GeV^2$).
The effect of anisotropy on D.E. of J/$\psi$ with temperature has been shown in Fig. \ref{fig.4}, D.E. of J/$\psi$ increases with increasing temperature, and on increasing the anisotropy from ($\xi$= -0.5 to 0.5) at fix value of magnetic field, D.E. decreases. From Fig. \ref{fig.4} we observe that D.E. become negative when anisotropy changes from -0.5 to -0.3 and become more and more negative on increasing the anisotropy.\\
The primary objective of this study is to determine the dissociation temperature from the intersection point of the binding energy and dissociation energy curves. We compared our results with previous findings and found a good degree of agreement.
Fig. \ref{fig.5} shows the variation of B.E. and D.E. of J/$\psi$ (left panel) and $\Upsilon$ (right panel) with T/$T_c$ for different values of anisotropy ($\xi$= 0.3, 0.5, 0.7) at fixed value of temperature(T= 0.3 GeV) and magnetic field (eB= 0.3 $GeV^2$). From Fig. \ref{fig.5} We observe that as anisotropy increases, B.E. increases while D.E. decreases.
Fig. \ref{fig.6} also shows the variation of B.E. and D.E. for $\psi$' (left panel) and for $\Upsilon$' (right panel) with T/$T_c$ for different values of anisotropy ($\xi$= 0.3, 0.5, 0.7) at fixed value of temperature(T= 0.3 GeV) and magnetic field (eB= 0.3 $GeV^2$).
From table-\ref{T1} it is noted dissociation temperature increases with anisotropy. From Fig. \ref{fig.5} and Fig. \ref{fig.6} we observed that if we increase the value of anisotropy, the dissociation temperature will also increases. The values of dissociation temperature for J/$\psi$, $\Upsilon$, $\psi$', $\Upsilon$' for different values of anisotropy at constant magnetic field can be seen in table-\ref{T1}.
Similarly, Fig. \ref{fig.7} and Fig. \ref{fig.8} also shows the variation of B.E. and D.E. in left panel for (J/$\psi$ and $\psi$') in right panel for ($\Upsilon$ and $\Upsilon$') for three different values of magnetic field (eB= 0.1, 0.2, 0.3 $GeV^2$) at fix value of anisotropy ($\xi$=0.2) and temperature (T= 0.3 GeV). From Fig. \ref{fig.7} and Fig. \ref{fig.8} we have seen that B.E. decreases with increasing magnetic field and D.E. increase with magnetic field.

In conclusion, from Figures \ref{fig.5}, \ref{fig.6}, \ref{fig.7}, \ref{fig.8} we determine that the dissociation temperature increases with anisotropy, while it decreases with increasing magnetic field. The values of dissociation temperature for  J/$\psi$, $\Upsilon$, $\psi$', $\Upsilon$' for different values of magnetic field at fix value of anisotropy can be seen in table-\ref{T2}.

\section{\label{sec:level4}Conclusions}
This study focuses on the effects of magnetic fields on the properties of heavy quarkonium in anisotropic media, utilizing a medium-modified Cornell potential that incorporates both perturbative and non-perturbative terms. 
The inclusion of the string term ($\sigma$) makes the potential more attractive, indicating that heavy quarkonium states become more bound when both the string and Coulombic terms are modified together. We have employed the real part of heavy quark potential to obtain the binding energy (B.E) and dissociation energy (D.E.). which in turn helps us to obtain the dissociation temperature of quarkonium states. It is observed that the B.E. has higher values with anisotropy. Similarly, the dissociation temperature of quarkonium states is found to be higher with anisotropy in comparison to the values obtained with the magnetic field. It is pertinent to mention here that the dissociation temperature obtained at $eB=0.3 GeV^2$ and $\xi=0.3$ in our model setup ranges in between the upper and lower limit of dissociation temperature obtained in ref.\cite{Lal_2023}.\\
To reinforce our findings, we compare them with holographic QCD models, which connect gauge theories with gravitational theories, offering deeper insights into quarkonia in strongly coupled regimes. As indicated in ref.\cite{Fadafan_2024}, an increase in the anisotropic parameter results in a decline in potential energy. Lower potential energy correlates with greater stability, suggesting a higher dissociation temperature for the quarkonium system. Therefore, we conclude that increasing anisotropy leads to an elevated dissociation temperature. Additionally, our results show that the dissociation temperature decreases with higher magnetic fields, which is consistent with recent literature \cite{Zhou_2020,Chen_2022}. Overall, this study enhances our understanding of quarkonium dynamics and the significant effects of anisotropy and magnetic fields on its stability and dissociation characteristics.

In the near future, we can explore the effects of non-uniform and time-dependent magnetic fields on quarkonium binding energy and dissociation temperatures. Investigating these dynamics will deepen our understanding of quarkonium behavior under extreme conditions, particularly in scenarios like heavy-ion collisions. In future we will extend our work to investigate the survival probability of various quarkonium states at different center-of-mass energies $\sqrt{S_{NN}}$ using parameters such as anisotropy, baryonic chemical potential, and magnetic fields. The results from this work may also aid in exploring extremely dense objects like neutron stars, as studied by the Compressed Baryonic Matter (CBM) experiment at the Facility for Antiproton and Ion Research (FAIR).


\section{References}
\markright{Bibliography}
{}

\end{document}